\documentclass[twoside,11pt]{article}
\usepackage{epsfig,rotate}

\def\Journal#1#2#3#4{{#1} {#2} (#4) #3 }

\def\PRL{\em Phys. Rev. Lett.}

\def\PREP{\em Phys. Rep.}

\def\PRD{{\em Phys. Rev.} D}
\def\PRC{{\em Phys. Rev.} C}

\def\RMP{{\em Rev. Mod. Phys.}}

\newcommand{\be}{\begin{equation}}
\newcommand{\ee}{\end{equation}}
\newcommand{\bea}{\begin{eqnarray}}
\newcommand{\eea}{\end{eqnarray}}

\begin{document}

\title{
Future galactic supernova neutrino signal: What can we learn?
\footnote{Talk at the School `Neutrinos in Astro, Particle
and Nuclear Physics', Erice, September 18-26, 2001.} 
}

\author{P. Vogel
\\
Physics Department, Caltech, Pasadena, CA 91125, USA
}

\maketitle

\begin{abstract}
The next supernova in our galaxy will be detected by a variety of neutrino detectors.
In this lecture I discuss the  set of observables needed to 
constrain the models
of supernova neutrino emission. They are the flux normalizations, and 
average energies, of each of the three expected components of the neutrino flux:
$\nu_e$, $\bar{\nu}_e$, and $\nu_x$ (all the other four flavors combined).
I show how the existing, or soon to be operational, neutrino detectors
will be able to determine the magnitude of these observables, and
estimate the corresponding rates. 
\end{abstract}
%\eject
%\tableofcontents
\section{Introduction}
When neutrinos from supernova 1987A were detected by the Kamiokande
\cite{kam87}
and IMB \cite{imb87}
collaborations, a new era of neutrino astrophysics began. Despite the
limited statistics (11 events in Kamiokande and 8 events in IMB)
the observation confirmed that the core collapse supernovae emit
most of their binding energy (a few$\times 10^{53}$ erg) in neutrinos,
that the duration of the neutrino emission is $\sim$ 10 seconds,
and that the average energy of the neutrinos (at least of the
$\bar{\nu}_e$, which were the only flavor actually seen) is $\sim$ 15 MeV, 
close to expectations. The observation of SN1987A lead to a flood of papers
analyzing its consequences
(for a relatively early review, see e.g. \cite{Schramm90}),
which is only very slowly diminishing with time. 

Historically, there were
seven supernovae in our galaxy proper recorded in the past thousand years,
and none in the last three centuries
(some were
not core-collapse SN which emit neutrinos, though). 
All of them were relatively close
to the solar system, so it is difficult to estimate the true
rate averaged over the whole galaxy from this record. Consensus
estimate of  core-collapse supernova rate in our galaxy is about
three times per century \cite{snrate}. Thus, the next 
Galactic supernova neutrino burst can come
at any time, tomorrow or in several decades. It is likely that
neutrinos from such supernova
will be detected by a variety of detectors, with much better statistics
than for SN1987A. Thus a wealth of new information is expected from
such unique event
which cannot be repeated in the productive lifetime of an average
physicist.
(Unfortunately, the present or planned neutrino detectors are unable
to observe supernovae in even the nearest galaxy, Andromeda, about
700 kpc away.)
Here I discuss some of the lessons that should, and hopefully will,
be extracted from the neutrino signal of the next supernova in our galaxy. 

There are several  areas of physics that will 
greatly benefit from
the supernova neutrino observations. They can be  divided
into three broad categories:
\begin{enumerate}
\item) Neutrino properties; mass, mixing, decay, etc. In particular,
one could use the time-of-flight
of the neutral current signal 
(dominantly $\nu_{\mu}$ and $\nu_{\tau}$) to reach
sensitivity to masses of about 30 eV for these neutrinos 
\cite{BVI,BVII}. This would represent an improvement by more than three
orders of magnitude for the mass associated with $\nu_{\mu}$, and by
almost six orders of magnitude  for the mass associated with $\nu_{\tau}$
when compared to the present direct neutrino mass limits \cite{PDG}. 
If, moreover, the neutrino emission is abruptly truncated by the collapse
of the proto-neutron star into a black hole, one can use this sharp cut-off
in the neutrino signal
to improve the time-of-flight sensitivity to masses of $\sim$ 6 eV for
$\nu_{\mu}$ and $\nu_{\tau}$ and to $\sim$1.8 eV for $\nu_e$ \cite{Beacom}.
\item) Supernova properties. From the neutrino signal it might be possible
to determine the luminosities and average energies of all three components
of the neutrino flux:  $\nu_e$,  $\bar{\nu}_e$, and $\nu_x$ 
(this notation will be
used from now on collectively for $\nu_{\mu}$, $\nu_{\tau}$ and their
antiparticles).
\item) Supernova localization. Using the angular distribution of the products
of the neutrino induced signal, or the timing of the signal
recorded in widely separated detectors, it might be
possible to find the direction towards the supernova independently, or prior to,
of the optical signal (for the discussion of this item, see \cite{BVIII}).
\end{enumerate}

I refer to the listed references regarding the items 1.) and 3.) and in the following
I will concentrate on the item 2.) - the determination of  supernova properties
from the neutrino signal. My aim is going to be a definition of a `template', i.e., a recipe
how to determine the required quantities and what signal and statistical accuracy
one may expect using the existing, or soon to be operational, detectors.
Substantial deviations from this template will mean either that the supernova behaves
in an unexpected way, or that neutrino oscillations affect the signal.
Obviously, general analysis of all possibilities is impossible before
the fact. However, the existence of such a template might help in preparing the
detectors for the supernova signal, particularly those like SuperKamiokande, SNO,
KamLAND or Borexino, which are built for a different purpose.

I will consider a `standard' supernova
(for a review of Type-II supernova theory see \cite{Bethe}), 
approximately at the center of the galaxy,
at the distance from Earth of 10 kpc. The binding energy, which is essentially fully
emitted in neutrinos, is assumed to be $3 \times 10^{53}$ ergs. It is easy
to understand the magnitude of the binding energy $E_B$ by using the simple estimate 
\be
E_B \simeq \frac{3}{5} \frac{G_N M^2}{R}, ~~{\rm where} ~R = 10~ {\rm km}, 
~~M \simeq 1.4 M_{\odot} ~.
\ee

Neutrinos are trapped in the hot and dense protoneutron star. The mean free path
of neutrinos,
\be
\lambda = \frac{1}{\rho \sigma} \sim 10 ~{\rm m ~ for} ~ 
\rho \sim 10^{38} ~{\rm nucleons/cm}^3 ~, ~ \sigma \sim 10^{-41} ~{\rm cm}^2 ~
\ee
is substantially shorter than the radius of the protoneutron star. In fact, the
trapping occurs already when the star radius is $\sim$ 100 km and the mean free
path becomes comparable to the scale height $h$ of the infalling matter
($h = kT/M_p g$ where $g \sim 10^{12}$ ms$^{-2}$ is the gravitational acceleration
at that radius).

Trapped neutrinos diffuse through the  protoneutron star. They leave
the star when they reach the so-called neutrinosphere,
essentially the radius where their mean free path is comparable to the corresponding
scale height at that point. Again, a crude estimate
of the diffusion time is just the product of the time duration
between successive scatterings and the number of steps,
\be
\tau_{diff} \sim \frac{\lambda}{c} \frac{R^2}{\lambda^2} ~\sim~ 10 ~{\rm s}~.
\ee

Throughout the star, neutrinos of all flavors are in equilibrium, with decreasing
temperature at increasing radii. Thus, the temperature of the outgoing 
neutrino flux for each flavor will be the characteristic temperature of the
corresponding neutrinosphere. Since the mean free paths of the different neutrino
flavors are different, the position of their neutrinospheres, and hence 
also the decoupling
temperatures, will be different as well. The $\nu_x$ neutrinos 
undergo only neutral current
interactions, hence their mean free path is longest, and thus their decoupling temperature
will be highest. Both $\nu_e$ and $\bar{\nu}_e$ have in addition also charged current
interactions. Moreover, since the star contains many more neutrons than protons, the
$\nu_e$ mean free path will be shorter 
(since $\nu_e$ interact with neutrons)  than the  $\bar{\nu}_e$  mean free path
(since $\bar{\nu}_e$ interact with protons). Hence
a hierarchy of decoupling temperatures (or mean energies) is expected,
\be
T(\nu_x) (\sim 8 {\rm MeV}) >
T(\bar{\nu}_e) (\sim 5 {\rm MeV})  > T(\nu_e)  (\sim 3.5 {\rm MeV}) ~,
\ee
or
\be
\langle E_{\nu_x} \rangle \sim 25 ~{\rm MeV} > 
\langle E_{\bar{\nu}_e} \rangle \sim 16 ~{\rm MeV} >
\langle E_{\nu_e} \rangle \sim 11 ~{\rm MeV} ~. 
\ee

At the same time, one expects that the total luminosity
will be equally shared by all neutrino flavors, so averaged over 
time
\be
\langle L_{\nu} \rangle 
\simeq \frac{E_B}{6 \tau_{diff}} \simeq 5\times10^{51} ~{\rm erg/s} ~
~{\rm for~all~6~flavors}~ .
\ee 
Note that the the short initial $\nu_e$ neutronization pulse has only
small luminosity when compared to $\langle L_{\nu} \rangle$ and is
going to be difficult to observe.
For a detailed description of the supernova neutrino emission, 
including the justification of the choice of the decoupling temperatures,
see Refs. \cite{Woosley,Janka}.

For each neutrino flavor the corresponding time averaged
flux at Earth will be therefore
\be
\langle f_{\nu} \rangle = \frac{2.6 \times 10 ^{11}}
{ \langle E_{\nu} \rangle {\rm (MeV)}} {\rm  ~ cm^{-2} ~s^{-1}} ~,
\ee
for the assumed 10 seconds emission time. With such a flux
and a typical cross section of $\sim 10^{-41}$ cm$^2$, one expects
few hundred charged current interactions with protons in 1 kton of water,
and few tens of events in 1 kton of iron (or other heavy target).
Clearly, very large detectors, operating for a long time,
are needed.

Thus the challenge for supernova neutrino observers is to detect
separately the three expected components of the neutrino flux:
the $\nu_e$ component through the charged reaction on bound neutrons
(i.e., on nuclei), the $\bar{\nu}_e$ component most easily
through the charged current reaction on free protons, and the
$\nu_x$ component through neutral current reactions.
For each of these components one should determine, ideally, not only the total rate,
proportional to $L_{\nu}/\langle E_{\nu} \rangle
\int \sigma(E_{\nu}) f(E_{\nu}) d E_{\nu} $,
($ f(E_{\nu})$ is the normalized energy distribution; typically assumed
to be the Fermi-Dirac thermal one)
but also the temperature, or equivalently $\langle E_{\nu} \rangle$.
If, and only if this program can be accomplished, can one reach
reliable conclusions about supernova astrophysics and/or neutrino
oscillations. 

\section{Detecting $\bar{\nu}_e$ and $\nu_e$ through charged current reactions}

It is relatively easy to detect $\bar{\nu}_e$, since most detectors contain free protons
and one can utilize the reaction $\bar{\nu}_e + p \rightarrow e^+ + n$ with 
large cross section and a characteristic signature of the time and position correlated
positron and neutron.

The cross section is well known. Neglecting the small neutron recoil energy 
($\sim E_{\nu}^2/M_p)$, one can 
simply relate the  positron energy to the incoming neutrino energy, 
\be
E_e^{(0)} = E_\nu - \Delta, ~~ \Delta = M_n - M_p = 1.293 ~{\rm MeV} ~.
\ee
The differential cross section to this ($M_p \rightarrow \infty$) order is
\be
\left(\frac{{\rm d}\sigma}{{\rm d cos}\theta}\right)^{(0)} =
\frac{\sigma_0}{2}
\left[(f^2 + 3g^2) + (f^2 - g^2) v_e^{(0)} \cos\theta \right]
E_e^{(0)} p_e^{(0)} ~, ~ \sigma_0 = \frac{G_F^2 \cos^2\theta_C}{\pi}
(1 + \Delta_{inner}^{R})
\ee
where the vector and axial-vector coupling constants are $f =
1$, $g = 1.26$ and $ \Delta_{inner}^{R} \simeq 0.024$
represents the inner radiative corrections. 
Integrating over angles one obtains the standard result
for the total cross section, which can be 
also related to the neutron lifetime $\tau_n$,
\be
\sigma^{(0)}_{tot}  = 
\sigma_0\; (f^2 + 3 g^2)\; E_e^{(0)} p_e^{(0)} 
 =  0.0952 \left(\frac{E_e^{(0)} p_e^{(0)}}{1 {\rm\ MeV}^2}\right)
\times 10^{-42} {\rm\ cm}^2 = \frac{2 \pi^2/m_e^5}{f^R_{p.s.} \tau_n}\; E_e^{(0)} p_e^{(0)}\,,
\ee
where  $f^R_{p.s.} =
1.7152$ is the phase space factor, including the Coulomb, weak
magnetism, recoil, and outer radiative corrections.
For supernova $\bar{\nu}_e$ terms of order $1/M_p$ should be included. The expressions for
the cross section to that order, 
including angular distribution, can be found in Ref. \cite{cross}.

For the `standard' SN with $T_{\bar{\nu}_e}$ = 5 MeV
one expects $\sim$8300 $e^+$ events in Superkamiokande,
and  $\sim$360 events in the light water part of SNO. SNO will be able to detect  
 $\bar{\nu}_e$ also by the charged current reaction on deuterons; one expects
about  $\sim$80 events of this kind with two neutrons in the final state. In KamLAND, which
is a scintillation detector, the correlation between the positron and the
neutron capture $\gamma$-rays can be used; one expects  $\sim$330 events there.
Altogether, it should be possible to measure with good accuracy the
luminosity and energy distribution of supernova   $\bar{\nu}_e$. In 
Superkamiokande the statistics ought to be sufficient to determine also
the  time dependence of the $\bar{\nu}_e$ luminosity and temperature.

It is more difficult to detect $\nu_e$ since they interact only with neutrons
and are expected to have lower temperature ( $T_{\nu_e}$ = 3.5 MeV). Both
$^{16}$O (in water \v{C}erenkov detectors) and $^{12}$C (in scintillation
detectors) have high thresholds for the $\nu_e$ induced charged 
current reactions,
15.42 MeV and 17.34 MeV, respectively. Thus, one expects negligible yields
for the charged current reactions on these targets as long as  $T_{\nu_e}$
is indeed only 3.5 MeV.

In SNO the `solar' reaction $\nu_e + d \rightarrow e^- + p + p$ with mere 1.44 MeV
threshold should yield about 80 events, perhaps sufficient to determine, at least
crudely, the temperature $T_{\nu_e}$ and the corresponding luminosity.

Since cross sections for the charged current $\nu_e$ interaction with nuclei
typically increase quickly with $E_{\nu}$, the count rates would increase dramatically
if $\nu_e \leftrightarrow \nu_x$ mixing occurs, which is likely
to happen. Hence observation of the $\nu_e$ signal represents a sensitive test
for oscillations. In KamLAND one expects only $\sim$2 events for the
$\nu_e$$^{12}$C $\rightarrow ^{12}$N$_{gs} e^-$ reaction if $T_{\nu_e}$ = 3.5 MeV.
That rate increases to $\sim$ 15 for vacuum oscillations, and to 27 for
the resonant MSW oscillations. (This reaction has an excellent signature since
one can use the delayed coincidence with the $^{12}$N $\beta^+$ decay.)
In SuperKamiokande the reaction 
$\nu_e~ ^{16}{\rm O} \rightarrow ^{16}$F$^* e^-$
results in only $\sim$ 20 events for $T_{\nu_e}$ = 3.5 MeV. However,
if through oscillations the effective $T_{\nu_e}$ = 8 MeV, the yield increases
dramatically to $\sim$860 events \cite{haxton,kolbe1}. The electrons from 
the $\nu_e$ charged current reaction on $^{16}$O can be distinguished,
in principle,
from the positrons from $\bar{\nu}_e$ on protons by their angular distribution.

Lead has been proposed as the target material in OMNIS and LAND supernova neutrino
detectors. The charged current reaction on $^{208}$Pb induced by $\nu_e$ has
threshold of only 2.9 MeV, but the corresponding strength is dominated by
the excitation of the giant Gamow-Teller resonance at about 16 MeV excitation
energy. The proposed detectors would register neutrons emitted by the
decay of the final $^{208}$Bi for the charged current reaction or $^{208}$Pb$^*$
for the neutral current reaction. Generally, it would be
difficult to separate the charged and neutral current responses
with such scheme. (Although with the `normal' hierarchy, Eqs. (4,5),
the neutral current signal would dominate.) However, in Ref. \cite{haxton2} 
it was shown that
the observation of the double neutrons could serve as a signature of the
charged current induced events in the case of oscillations. One drawback
is that the corresponding cross sections for both reactions are rather
uncertain. In fact, the two recent calculations of these
quantities \cite{haxton2,kolbe2} differ by about a factor of two. Thus, if
the lead based supernova detectors are ever build, experimental determination 
of these cross sections will be necessary.

\section{Detecting  $\nu_x$ neutrinos through  neutral current scattering}

The supernova $\nu_x$, i.e. $\nu_{\mu}$ and $\nu_{\tau}$ with their antiparticles,
do not have enough energy to induce charged current (CC) interactions. Thus, they can
be detected only through their neutral current (NC) scattering. In order to detect
the NC scattering one has to find, first of all, the appropriate signature, i.e. a
reaction that can be clearly recognized and separated from the CC channels. Since NC
scattering is flavor blind, the contribution of the $\nu_e$ and $\bar{\nu}_e$ scattering
has to be subtracted in order to isolate the $\nu_x$ effect. This condition
more or less eliminates neutrino-electron scattering, where the $\nu_e$ and $\bar{\nu}_e$
contribution dominates. However, in semileptonic NC scattering the cross section
typically increases fast with energy, and hence the $\nu_x$ contribution will
dominate the NC yield. (The fact that there are four flavors in the $\nu_x$
flux helps as well.)

The other difficulty is that in a typical NC reaction there is no spectral information;
only the number of events per unit time can be measured. Generally, 
the scattering rate (per s) is: 
\be
\frac{dN_{NC}}{dt} = C
\int dE\,f(E) \left[\frac{\sigma(E)}{10^{-42} {\rm cm}^2}\right]
\left[\frac{L(t - \Delta t(E))}{E_B/6}\right]\,,
\label{eq:ncrate}
\ee
where for SuperKamiokande
\be
C = 9.21
\left[\frac{E_B}{10^{53} {\rm\ ergs}}\right]
\left[\frac{1 {\rm\ MeV}}{T}\right]
\left[\frac{10 {\rm\ kpc}}{D}\right]^2
\left[\frac{{\rm det.\ mass}}{1 {\rm\ kton}}\right]
\,n\,,
\label{eq:C}
\ee
$T$ is the spectrum temperature (where we assume $\langle E \rangle =
3.15 T$, as appropriate for a Fermi-Dirac spectrum), $f(E)$,
the neutrino energy distribution is in MeV$^{-1}$,
and $n$ is the number of target nuclei per water molecule. 
Also, $ \Delta t(E)$
is the possible delay caused by the finite neutrino mass.

Thus, NC scattering rate depends on both the luminosity and temperature,
and their effects cannot be directly separated.
On the other hand, the NC signal is obviously independent of possible oscillations
between active (as opposed to sterile) neutrinos.

\begin{figure}[htb]
\begin{center}
\epsfig{file=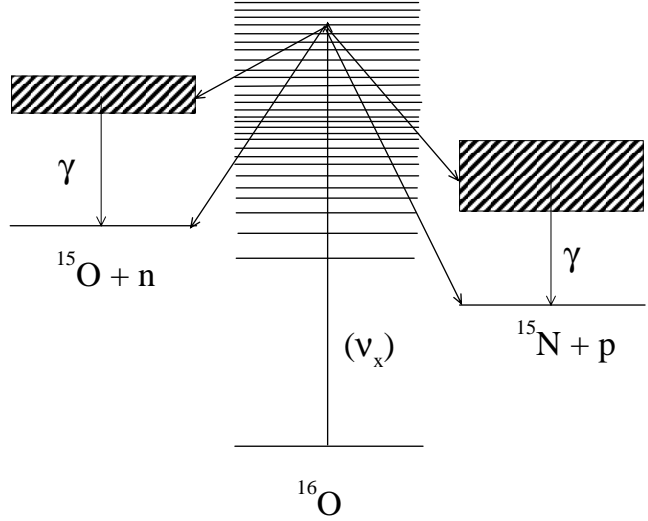,scale=0.7}
\end{center}
\caption{Schematic illustration of the detection scheme for supernova
$\nu_x$ neutrinos in water \v{C}erenkov detectors.\label{fig:o16sn}}
\end{figure}

There are several ways in which the NC signal in existing detectors can be
determined. In water \v{C}erenkov detectors, one can use the
signal proposed in \cite{LKV} according to which the $\nu_x$ 
neutrinos will excite $^{16}$O
into the continuum that will deexite dominantly by the emission of either proton
or neutron. There is a sizable probability (about 30\%) that the
resulting $^{15}$N or $^{15}$O nucleus will be in a bound excited
state, as indicated in Fig. \ref{fig:o16sn}. These states, in turn, deexite
by $\gamma$ emission with characteristic energies between 5 and 10 MeV,
above the SuperKamiokande threshold, and
easily separated from the background positrons from $\bar{\nu_e} p \rightarrow e^+ n$.
In SuperKamiokande one expects about 700 events of this kind.

\begin{figure}[h!!]
\begin{center}
\rotate[r]{\epsfxsize=0.45\textwidth
\epsffile{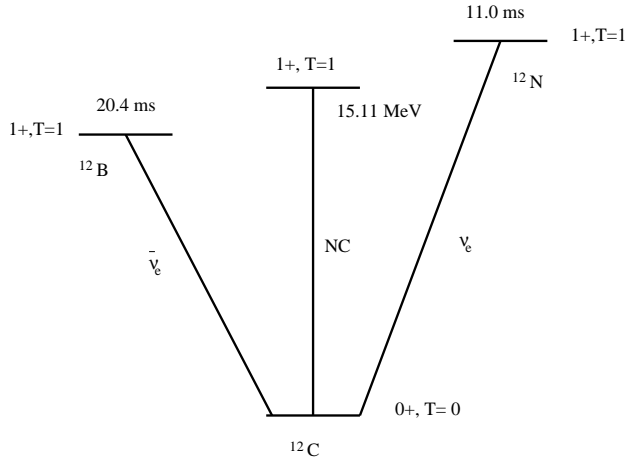}}
\end{center}
\caption{Illustration of the excitation  of the
$T=1, ~I=1^+$ triad in mass $A = 12$ nuclei .\label{fig:c12triad}}
\end{figure}

In SNO the obvious NC signal is the deuteron disintegration,
$\nu_x + d \rightarrow n + p + \nu_x$. It can be recognized by 
detecting a single neutron, but no electron (or positron).
The rate is obtained by the same equations as above, except in 
Eq. (\ref{eq:C}) one should replace 9.21 $\rightarrow$ 8.28.
There will about 400 NC $\nu_x$ induced events in SNO, with
another $\sim 85$ induced by $\nu_e$ and $\bar{\nu}_e$.

In scintillation detectors, such as KamLAND or Borexino, $\nu_x$ NC
scattering with excitation of the 15.11 MeV $T = 1, I^{\pi} = 1^+$
state in $^{12}$C is possible
(see Fig. \ref{fig:c12triad}).
This process offers a very distinct signature and has the further advantage
that the corresponding cross section is calculable accurately, and has
been verified by the KARMEN and LSND experiments. One expects $\sim$ 60
events with 15.11 MeV $\gamma$ in KamLAND.

Finally, as mentioned above, in lead based detectors the single neutron events 
will be dominated by the NC $\nu_x$ scattering.

Thus, there will be a rather accurate information on the rate of the NC events. By
combining the data from different detectors, one can try to determine
the $\nu_x$ luminosity and temperature separately. 
This should be possible, at least crudely,
since the mentioned reactions, while all proportional to the $\nu_x$ luminosity,
will have slightly different dependence on neutrino energy in the various respective
cross sections. 

\section{Neutrino elastic scattering on protons}

Ideally, one would like to use NC scattering combined with some spectrum information,
not just rate as in the previous section. As stressed previously also, the seemingly obvious
candidate process, neutrino - electron scattering, will be dominated by the $\nu_e$ and
$\bar{\nu}_e$ scattering, and thus is not very convenient to study the $\nu_x$ scattering.

\begin{figure}[ht]
\begin{center}
\epsfig{file=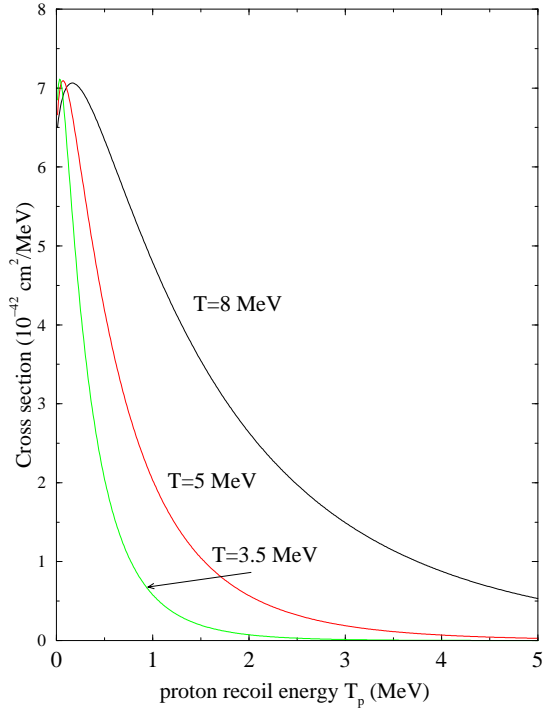,scale=0.45}
\end{center}
\caption{Cross section of the elastic neutrino scattering on protons for
the indicated temperatures of the incoming neutrinos. Proton
recoil energies without quenching are used. \label{fig:nup}}
\end{figure}

In detectors with low detection threshold, such as the scintillator based KamLAND
and Borexino, one can, in principle use 
for this purpose the elastic scattering on protons
\footnote{The content of this section is based on the suggestion of
John Beacom, for details see \protect\cite{will}.}.
The corresponding differential cross section is 
\be
\frac{{\rm d} \sigma}{{\rm d} T_p} =  \frac{G_F^2 M_p}{\pi}
\left[(c_A^2 + c_V^2) - (c_A^2 - c_V^2)\frac{T_p M_p}{2E_{\nu}^2}
 -  (c_V \mp c_A)^2\frac{T_p}{E_{\nu}} \pm 2c_Mc_A\frac{T_p}{E_{\nu}}
\right] ~,
\ee
where $c_V = 1/2 - 2 \sin^2 \theta_W = 0.0375, ~c_A = 1.26/2$,
$c_M \simeq -\mu_n/2$, and $\pm$ refers to $\nu$ and $\bar{\nu}$, respectively.
(We have neglected the possible effect of the strangeness
component of the proton). The total cross section is 
proportional to $E_{\nu}^2$, so the signal
will be dominated by $\nu_x$, particularly above
reasonable detection thresholds.
However, while the recoiling protons  scintillate, 
the scintillation light is quenched, compared
to electrons or $\gamma$. Thus, the relevant 
observable energies are $\le$ 1 MeV, and difficult
to detect and separate from backgrounds. However, in a sensitive 
low background detector 
one might be able not only to count the
number of events, but actually observe the proton recoil spectrum.
The cross section, without account of quenching, and for the Fermi-Dirac
spectrum of incoming neutrinos, is shown in Fig. \ref{fig:nup}. Note the
sharp dependence on the neutrino temperatures.

\begin{figure}[h!!!]
\begin{center}
\epsfig{file=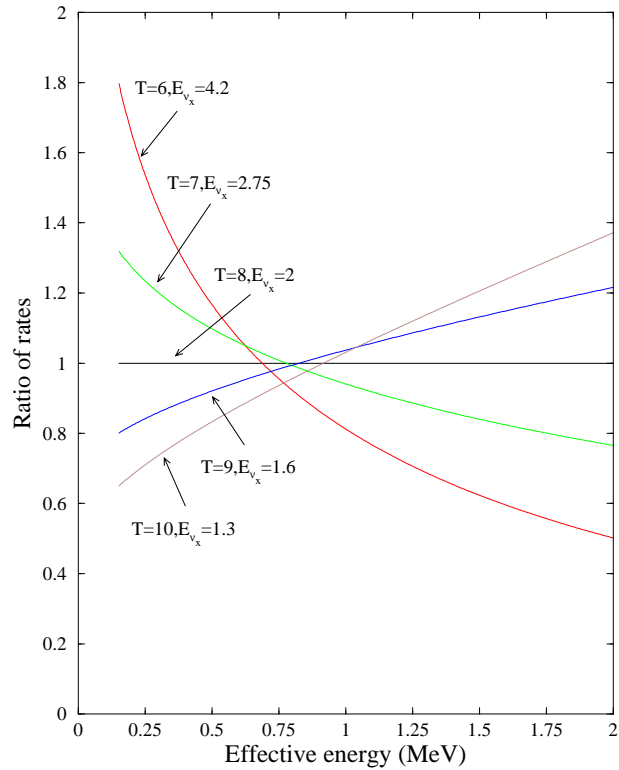,scale=0.5}
\end{center}
\caption{Ratio of proton yields, as a function of the effective quenched
energy, to the standard case of $T = 8$ Mev, and the total energy emitted
in $\nu_x$ equal to 2$\times 10^{53}$ erg. All considered cases result
in the same total number of events above the threshold of 200 keV
of the effective energy.\label{fig:ratios}}
\end{figure}

\begin{figure}[h!!!]
\begin{center}
\epsfig{file=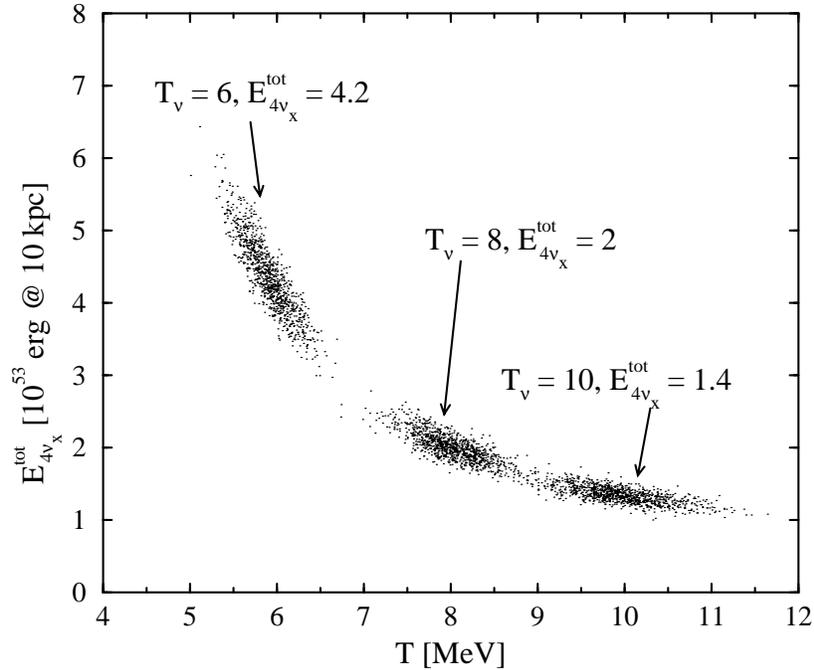,scale=0.65}
\end{center}
\caption{Monte Carlo simulation of the combined fit to $T_{\nu_x}$ and the total
energy carried by such neutrinos, $E_{4\nu_x}^{tot}$.\label{fig:will}}
\end{figure}

Let us assume that one will be able to extract from measurement some spectral information
on the recoiling protons. Would that make it possible to distinguish the
cases in which the $\nu_x$ luminosity and temperature conspire in such a way
that they lead to the same total number of events, 
and therefore are indistinguishable based only on Eq. (11)? The answer is yes, and how
this could be accomplished is illustrated in Fig. \ref{fig:ratios}. One can see that the
proton recoil spectra sensitively depend on the neutrino temperature, with the
ratio of the low and high energy yields decreasing with the increasing temperature.
In a detailed simulation \cite{will} the power of such discrimination was
demonstrated by taking into account the statistical fluctuation of the expected
data. As shown in Fig. \ref{fig:will} one expects about 10\% resolution
on both the $\nu_x$ temperature and total energy carried by these neutrinos.

It should be stressed once more that the considered NC signal is independent of neutrino
oscillations into `active' flavors, i.e. $\nu_e \leftrightarrow \nu_{\mu,\tau}$ and
obviously  $\nu_{\mu} \leftrightarrow \nu_{\tau}$. If this signal can be in fact
detected, it would measure the luminosity and temperature of the hottest
component of the supernova neutrino emission spectrum.

\section{Conclusions}

In this lecture I have shown how, through the combination of the existing
(or soon to be operational) detectors, one can determine simultaneously
and independently the luminosities and average energies (or temperatures)
of the three expected components, $\nu_e$, $\bar{\nu}_e$
and $\nu_x$, of the next Galactic supernova neutrino flux. For a `standard' supernova
near the center of our galaxy, at 10 kpc, I have estimated the corresponding
count rates, neglecting for a moment the possible effects of neutrino
oscillations.

Having this set of quantities will make it possible to verify, or find deviations,
from the basic assumptions about the supernova neutrino emission: the equal luminosity
in each of the six neutrino flavors, and the hierarchy of decoupling temperatures.
Also, one should be able to determine the total emitted energy, essentially
the supernova binding energy, and the total neutrino fluence. Such observables will,
in turn, severely constrain theoretical models of supernova neutrino emission,
and allow one to deduce conclusions about the possible role of neutrino 
oscillations.

Most of the original results reported here were obtained in 
a highly pleasurable collaboration with John Beacom. The work
was supported by the US DOE contract DE-FG03-88ER40397.

\end{document}